\PassOptionsToPackage{table}{xcolor}
\documentclass[sigconf,screen]{acmart}

\usepackage{comment}
\usepackage[utf8]{inputenc}
\usepackage{titlesec}
\usepackage[table]{xcolor}
\usepackage{geometry}
\usepackage{pifont}
\usepackage{array}
\usepackage{lipsum}
\usepackage{enumitem}
\usepackage{mathastext}

\usepackage{booktabs}
\usepackage{xcolor} 

\usepackage{lipsum}  
\usepackage[utf8]{inputenc}
\usepackage{xspace}
\usepackage{ragged2e} 
\usepackage{rotating} 
\usepackage{fontawesome5} 
\usepackage{caption} 

\definecolor{lightgray}{rgb}{0.95, 0.95, 0.95} 
\definecolor{darkgray}{rgb}{0.7, 0.7, 0.7} 
\definecolor{mygreen}{rgb}{0.0, 0.5, 0.0} 
\definecolor{myred}{rgb}{0.8, 0.0, 0.0} 

\newcolumntype{C}{>{\bfseries}c} 
\definecolor{main}{HTML}{cccccc}    
\definecolor{sub}{HTML}{000000}     

\AtBeginDocument{%
  }

\acmYear{2025}\copyrightyear{2025}
\setcopyright{rightsretained}
\acmConference[FSE Companion '25]{33rd ACM International Conference on the Foundations of Software Engineering}{June 23--28, 2025}{Trondheim, Norway}
\acmBooktitle{33rd ACM International Conference on the Foundations of Software Engineering (FSE Companion '25), June 23--28, 2025, Trondheim, Norway}
\acmDOI{10.1145/3696630.3728720}
\acmISBN{979-8-4007-1276-0/25/06}

\begin{document}

\title{A Path Less Traveled: Reimagining Software Engineering Automation via a Neurosymbolic Paradigm}

\author{Antonio Mastropaolo}
\affiliation{%
 \institution{William \& Mary}
 \city{Williamsburg}
 \state{Virginia}
 \country{USA}}
\email{amastropaolo@wm.edu}

\author{Denys Poshyvanyk}
\affiliation{%
  \institution{William \& Mary}
  \city{Williamsburg}
  \state{Virginia}
  \country{USA}}
\email{dposhyvanyk@wm.edu}

\definecolor{darkgreen}{rgb}{0.0, 0.5, 0.0}
\newcommand{\re}{\textcolor{red}{\textbf{[REF]}\xspace}}
\newcommand{\cmark}{\ding{51}}%
\newcommand{\xmark}{\ding{55}}
\newcommand{\rev}[1]{\textcolor{black}{#1}}
\newcommand{\ie}{\emph{i.e.,}\xspace}
\newcommand{\eg}{\emph{e.g.,}\xspace}
\newcommand{\etc}{etc.\xspace}
\newcommand{\etal}{\emph{et~al.}\xspace}
\newcommand{\secref}[1]{Section~\ref{#1}\xspace}
\newcommand{\chapref}[1]{Chapter~\ref{#1}\xspace}
\newcommand{\appref}[1]{Appendix~\ref{#1}\xspace}
\newcommand{\figref}[1]{Fig.~\ref{#1}\xspace}
\newcommand{\listref}[1]{Listing~\ref{#1}\xspace}
\newcommand{\tabref}[1]{Table~\ref{#1}\xspace}
\newcommand{\greenAI}{\emph{\textcolor{ForestGreen}{Green} AI}\xspace}
\newcommand*\circled[1]{\tikz[baseline=(char.base)]{
		\node[shape=circle,fill,inner sep=0.8pt] (char) {\textcolor{white}{#1}};}}
\newboolean{showcomments}
\setboolean{showcomments}{true}

\ifthenelse{\boolean{showcomments}}
{\newcommand{\nb}[2]{
		\fbox{\bfseries\sffamily\scriptsize#1}
		{\sf\small$\blacktriangleright$\textit{#2}$\blacktriangleleft$}
	}
	\newcommand{\cvsversion}{\emph{\scriptsize$-$Id: macro.tex,v 1.9 2005/12/09 22:38:33 giulio Exp $}}
}
{\newcommand{\nb}[2]{}
	\newcommand{\cvsversion}{}
}

\newcommand{\verylightgray}[1]{\cellcolor{gray!10}{#1}}
\newcommand{\lightgray}[1]{\cellcolor{gray!22}{#1}}
\newcommand{\gray}[1]{\cellcolor{gray!33}{#1}}
\newcommand{\darkgray}[1]{\cellcolor{gray!45}{#1}}

\newcommand\ANTONIO[1]{\textcolor{red}{\nb{ANTONIO}{#1}}}

\begin{abstract}
The emergence of Large Code Models (LCMs) has transformed software engineering (SE) automation, driving significant advancements in tasks such as code generation, source code documentation, code review, and bug fixing. However, these advancements come with trade-offs: achieving high performance often entails exponential computational costs, reduced interpretability, and an increasing dependence on data-intensive models with hundreds of billions of parameters. In this paper, we propose \emph{Neurosymbolic Software Engineering}, in short NSE, as a promising paradigm combining neural learning with symbolic (rule-based) reasoning, while strategically introducing a controlled source of \emph{chaos} to simulate the complex dynamics of real-world software systems. This hybrid methodology aims to enhance efficiency, reliability, and transparency in AI-driven software engineering, while introducing controlled randomness to adapt to evolving requirements, unpredictable system behaviors, and non-deterministic execution environments.  By redefining the core principles of AI-driven software engineering automation, NSE lays the groundwork for solutions that are more adaptable, transparent, and closely aligned with the evolving demands of modern software development practices.

\end{abstract}



\maketitle


\section{Introduction}
\label{sec:intro}

During the past decades, software engineering automation has undergone a transformative evolution, driven by the rise of deep neural networks (DNN) and the development of large-scale deep learning (DL) models with hundreds of billions of parameters, such as DeepSeek-CoderV2 \cite{deepseekv2}. The scalability of these models, achieved through extensive pre-training on diverse datasets and increasingly larger architectures, has been a cornerstone of their success. This has enabled large language models (LLMs) to develop a deep understanding of software engineering principles, automating tasks such as code generation, documentation, optimization, and bug fixing \cite{10.1145/3485275,hou2024large}. These advancements have revolutionized development workflows, significantly boosting the efficiency and accuracy of software creation and maintenance.

The impact of LLMs is further evidenced by the rapid adoption of commercial AI-powered tools such as GitHub Copilot \cite{copilot} and ChatGPT \cite{chatgpt}. For example, GitHub Copilot, Microsoft's flagship AI-driven solution for software engineering, surpassed one million active users by June 2023\footnote{\url{https://aibusiness.com/companies/one-year-on-github-copilot-adoption-soars}}, underscoring the increasing dependence on AI--assisted tools within the developer community. What was once a distant vision: the deep integration of AI into software engineering has now become a tangible reality, signaling a transformative shift in the field. AI-powered tools have transitioned from experimental prototypes to essential components of modern software engineering workflows.

This rapid progress has been fueled by two key factors: (i) the unprecedented accessibility of data, particularly from public code repositories like GitHub, and (ii) significant technological advancements in hardware acceleration, such as GPUs and TPUs. These developments have enabled the scaling of LLMs and large code models (LCMs) to hundreds of billions of parameters, empowering AI-driven coding assistants like GitHub Copilot \cite{copilot} and AWS CodeWhisperer \cite{codewhisperer} to deliver highly accurate, efficient, and context-aware solutions. As a result, AI-enabled software engineering has become not only feasible but also a cornerstone of modern development practices, reshaping the field in profound ways.




However, as these models become increasingly complex, the increasing demand for computational power results in higher energy consumption, higher hardware costs, and a greater environmental impact \cite{hou2024large}. The relentless drive toward larger architectures and expanding resource requirements not only strains existing infrastructure but also raises a fundamental question: \emph{Is indefinite scaling truly a sustainable and effective path for advancing software engineering automation?}

A possible answer lies in the work of Villalobos \etal \cite{villalobosposition}, anticipating the beginning of a new phase, where the volume of human-generated data can no longer sustain the exponential growth of LLMs. This imminent change, which we refer to with the term \textbf{singularity of automation} marks a turning point where AI’s demand for training data exceeds the production rate of human knowledge. 

In the context of software engineering, this phenomenon was first introduced and explored in the pioneering work of Velasco \etal \cite{velasco2025toward}, highlighting the urgent need for alternative approaches beyond the mere scaling of models. In particular, Velasco's work \etal challenged the conventional belief that scaling deep learning models alone is the most effective way to push the boundaries of AI-driven software engineering. In this context, the authors introduced the first Neurosymbolic framework specifically tailored for program comprehension tasks, leveraging the complementary strengths of deep learning techniques (\eg LLMs) and traditional symbolic methods to enhance program comprehension activities.

Building on this foundation, we introduce Neurosymbolic Software Engineering (NSE), a generalized extension of the Neurosymbolic Program Comprehension (NsPC) framework, expanding its scope beyond program comprehension to encompass the full spectrum of SE-related tasks. NSE offers a compelling alternative to the indiscriminate scaling of deep learning models, while preserving the key benefits that deep learning-based techniques have introduced to the field.

A key improvement of the novel paradigm is the integration of \emph{chaos}, a critical yet previously unexplored component in the original framework by Velasco \etal \cite{velasco2025toward}. This addition allows us to address a fundamental challenge in software engineering: modeling and managing uncertainty in complex, evolving systems.
To achieve this, we leverage the mathematical foundation of chaos theory \cite{gleick2008chaos}, a powerful framework that provides a structured approach to understanding unpredictability, shifting requirements, and dynamic behaviors in software development. Unlike pure randomness, which is entirely unpredictable, a chaotic system operates under deterministic rules but is highly sensitive to initial conditions. This means that even small changes in input can lead to drastically different outcomes over time, a phenomenon widely known as the ``butterfly effect''--where a minor variation in one part of a system can cascade into significant and unexpected consequences elsewhere.

In this position paper, we advocate for the integration of chaos theory into software engineering automation pipelines as an effective and efficient proxy to model ``true randomness''. 
With that in mind, we introduce a unified vision that merges symbolic reasoning and probabilistic learning with chaos, naturally bridging between these two domains. This hybrid approach seeks to improve transparency, efficiency, trustworthiness, and sustainability in AI-driven software engineering, ensuring that automation remains scalable, interpretable, and practically applicable while capturing the dynamic aspects of real-world software systems.
The remainder of this paper explores the key components of the NSE paradigm we envision, its advantages, and its broader implications for the future of AI in software engineering. 
\vspace{2pt}


\vspace{-12pt}
\section{Background and Related Work}
In this section, we first provide an overview of recent advances in LCMs and their impact on various software engineering tasks. We then highlight first attempts of Neurosymbolic approaches in SE. \\

\vspace{-15pt}
\subsection{Large Code Models}
Recent advancements in LCMs have substantially contributed to the automation of SE tasks, allowing greater support and greater accuracy in various SE-related processes \cite{hou2024large,10.1145/3485275}.
Although general-purpose LLMs, such as GPT-4 \cite{achiam2023gpt} and Claude \cite{Anthropic_Claude3}, exhibit broad applicability across multiple domains, they often lack the domain-specific granularity required for code-related tasks. To address this limitation, researchers have developed specialized LCMs, explicitly designed to enhance SE applications. Notable examples include CodeLlama \cite{roziere2023code}, StarCoder \cite{lozhkov2024starcoder}, Incoder \cite{fried:arxiv2022} and DeepSeek-Coder \cite{liu2024deepseek}. These domain-adapted models leverage code-centric training data and employ task-aware optimizations, thereby improving the precision and automation of SE activities, including various code-related tasks. A complete analysis is provided in the Systematic Literature Review by Watson \etal \cite{watson:tosem2022}. 

\subsection{Neurosymbolic AI in SE}

Princis \etal \cite{princis_sql_2024} introduce a hybrid approach that combines symbolic reasoning with LCMs to improve SQL query generation. Their system uses symbolic checks for query validation and repair, along with partial query evaluation and early elimination of invalid queries, resulting in improved runtime efficiency and accuracy.

Arakelyan \etal \cite{arakelyan_ns3_2022} enhance semantic code search (SCS) systems by integrating neural and symbolic methods. Their approach uses rule-based parsing to match natural language queries with code snippets, though the manually created rules may lack generalizability across languages.

In different research, Jana \etal \cite{jana_cotran_2024} introduces CoTran, a neurosymbolic system powered by large language models to translate code between programming languages. CoTran incorporates symbolic execution feedback to ensure the functional equivalence of translated code, specifically supporting translations between Java and Python. The inclusion of the symbolic component enhances the system's ability to preserve the original code's logic, resulting in more robust and reliable translations.

Parisotto \etal \cite{parisotto_neuro-symbolic_2016} propose a Neuro-Symbolic Program Synthesis approach to improve program induction, addressing key challenges such as high computational costs, limited interpretability, and task-specific training requirements. Their method constructs executable programs in a domain-specific language using input-output examples, demonstrating a strong generalization to unseen tasks.

In a separate line of research, Hu \etal \cite{hu_fix_2022} introduces NSEdit, a Transformer-based code repair method that predicts editing sequences to fix buggy source code across varied program structures and code repair benchmarks.

Further applications of neurosymbolic AI techniques can be found in the realms of representation learning \cite{allamanis_learning_2017}, error correction \cite{xue_interpretable_2024} and semantic code repair \cite{devlin_semantic_2017}.

\vspace{-5pt}
\section{Conceptualizing Neurosymbolic Software Engineering}

The rise of LCMs has closely driven the evolution of software engineering automation, enabling breakthroughs in tasks like code generation, bug fixing, and vulnerability detection--long viewed as beyond the reach of automation (\secref{sec:intro}). However, this progress has also revealed new challenges that now constrain further advances in the field.

This section introduces the core components of the proposed NSE paradigm and explains how their interaction supports AI-driven software automation (\secref{sec:nse}). We then highlight opportunities and open challenges for adopting NSE, outlining key directions for future research and practice (\secref{sec:opc}).

\subsection{Establishing a New Paradigm} 
\label{sec:nse}

We conceptualize \textbf{N}eurosymbolic \textbf{S}oftware \textbf{E}ngineering as an innovative paradigm to advance software engineering automation, harnessing the combined strengths of neural probabilistic learning and symbolic reasoning, unified through the presence of controlled chaos. The latter serves as the connective element between these two domains, enabling a synergistic collaboration that integrates the advantages of data-driven probabilistic learning, facilitating pattern recognition and deep semantic understanding, with the precision, interpretability, and correctness of symbolic reasoning, which has long been the foundation of rule-based frameworks and verification methods.

\begin{figure}[h]
\centering
\includegraphics[width=0.3\textwidth]{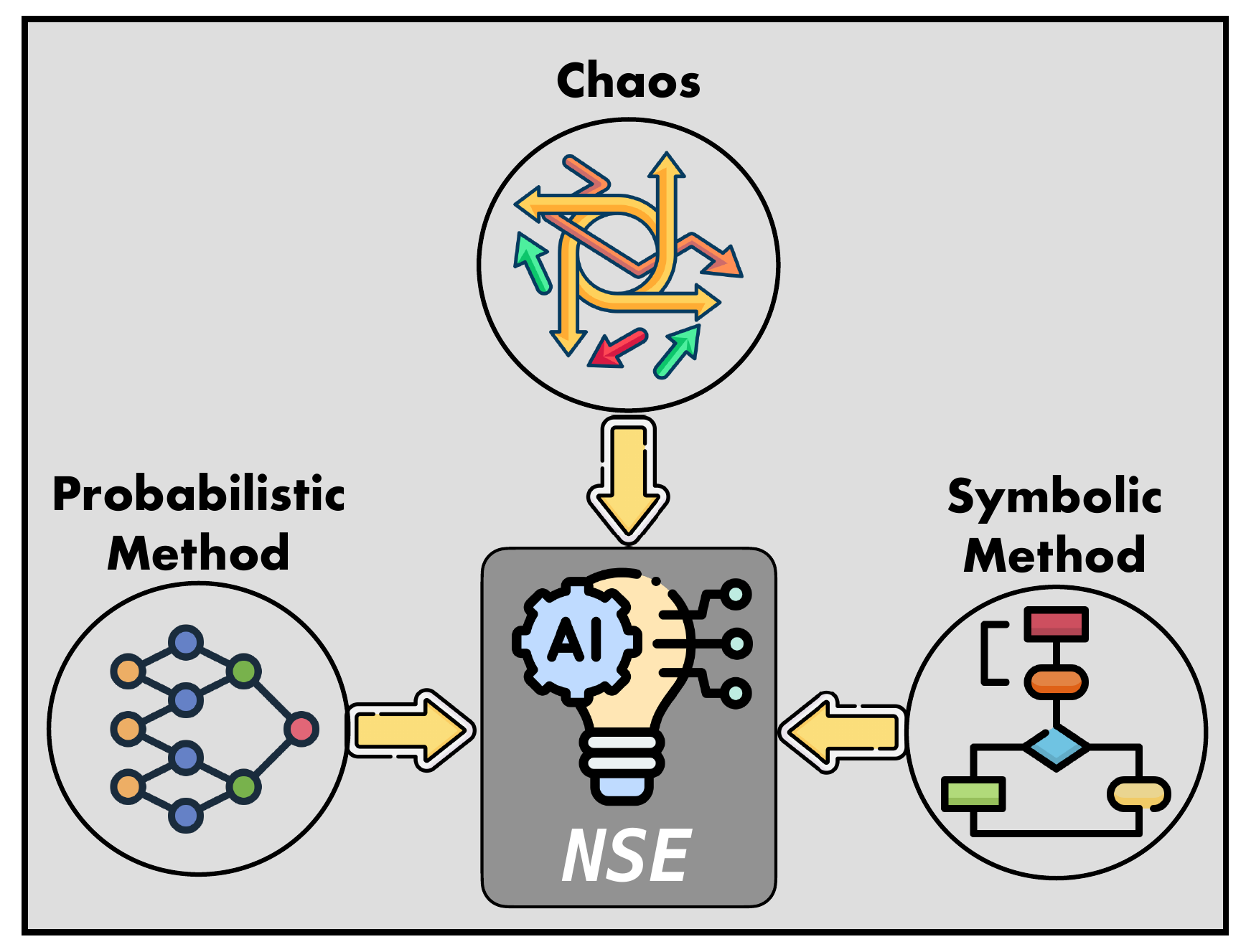}
\caption{Neurosymbolic Software Engineering Paradigm.}
\label{fig:nse-approach}
\end{figure}

Figure \ref{fig:nse-approach} illustrates the three fundamental components of our envisioned NSE paradigm. At its core, this paradigm integrates the following elements:

\begin{itemize}
\vspace{2pt}
\item \textbf{Probabilistic Method:} This component capitalizes on the latest advancement in LLMs and their ability to distill knowledge from vast amounts of code and natural language data. By identifying patterns and relationships within the data, LLMs and LMCs in particular enable systems to generalize and adapt to new unseen problems. These models excel at tasks such as code completion, bug detection, and natural language understanding, providing a data-driven foundation for automation in software engineering.

\vspace{5pt}
\item \textbf{Symbolic Method:} This component encompasses logical reasoning frameworks, constraint solvers, and rule-based inference systems. Symbolic methods ensure that software engineering processes are interpretable, correct, and aligned with formal specifications. By combining symbolic reasoning with probabilistic learning, the NSE paradigm achieves a balance between flexibility and precision.

\vspace{5pt}
\item \textbf{Chaos-driven Component:} \rev{This component injects controlled randomness into the system to simulate real-world uncertainty. Its primary function is to bridge neural probabilistic learning with symbolic reasoning, enabling the exploration of alternative solutions when deterministic logic or statistical inference alone falls short. During training, this component can be leveraged to introduce structured noise to simulate out-of-distribution data, situations where the input deviates from what the model has previously encountered. This ability to remain robust and effective despite such variations is a significant advantage, especially in dynamic software environments. Crucially, because the chaos-driven component employs structured randomness \ie variability governed by deterministic rules, it provides a principled approach to modeling uncertainty, allowing NSE to handle unpredictability with both flexibility and precision.}

\end{itemize}

Together, these three components form a cohesive framework that advances software engineering automation by combining the strengths of probabilistic learning, symbolic reasoning, and chaos-driven adaptability. This integration ensures that the resulting systems are not only powerful and efficient but also interpretable, trustworthy, and capable of addressing the complexities of real-world software engineering challenges.

\rev{In practice, one possible instantiation of NSE intentionally introduces controlled perturbations or noise into inputs and execution conditions during testing to observe how small changes propagate through the software pipeline. These chaos-driven trials can reveal hidden vulnerabilities and edge case behaviors that purely probabilistic or deterministic methods might overlook. Unlike random testing, which can lack focus, chaos in NSE is carefully managed and targeted, guided by theoretical principles to explore critical variations without introducing irrelevant or meaningless noise. This approach not only improves testing activities, but also provides valuable benefits for program repair, code generation, and security vulnerability detection in a broader context.}

\vspace{-10pt}
\subsection{Opportunities and Open Challenges}
\label{sec:opc}
By bridging neural learning and symbolic reasoning while incorporating deliberately controlled \emph{chaos}, NSE introduces a new class of automation solutions with unique advantages. The following sections provide an overview of the opportunities (\faLightbulb) that this paradigm offers in addressing key software engineering challenges.

Furthermore, we explore the open challenges (\faExclamationTriangle) that the SE community may face when transitioning to this approach. These challenges highlight critical areas that require further research, development, and innovation to fully exploit the potential of NSE and establish it as a transformative actor in the field, with the goal of fully integrating neurosymbolic principles into real-world software engineering workflows. \\

\vspace{-8pt}
\subsubsection{Opportunities for a better, faster, stronger AI-driven software engineering automation}
\scalebox{1.2}{\faLightbulb}  \\

\vspace{-8pt}
\textbf{Improved interpretability and scalability}: Unlike black-box deep learning models, NSE incorporates rule-based logic that allows developers to trace and comprehend the reasoning behind automated suggestions and predictions. Furthermore, the presence of the symbolic reasoner facilitates performance optimization via structured constraints, avoiding the necessity to scale up the size of the training dataset and DL architecture, to improve the performance of the system on the given task. \\

\vspace{-8pt}
\textbf{Trustworthiness and robustness against bias}: With the possibility to integrate formal verification thanks to the symbolic methods, NSE guarantees that the generated software engineering artifacts comply with well-defined correctness standards. Additionally, while traditional deep learning models are susceptible to adversarial perturbations, often resulting in misleading predictions, the integration of a symbolic reasoning component helps mitigate biases and robustness issues by enforcing logical consistency and rule-based validation mechanisms.\\

\vspace{-8pt}
\textbf{Energy-Efficient AI for Sustainable Automation}: 
Scaling deep learning models requires substantial computational resources, raising concerns about their environmental sustainability \cite{shi2024efficient}. NSE presents a more energy-efficient alternative by combining symbolic reasoning, structured constraints, and compact, specialized neural models. Specifically, NSE reduces the dependence on massive datasets by using structured symbolic knowledge, minimizing the need for continuous data expansion. In addition, as an extension of its scalability benefits, NSE enhances model sustainability by integrating predefined formal rules, logical constraints, and domain-specific knowledge, ultimately reducing the reliance on large-scale datasets for training while maintaining high performance and accuracy. \\

\vspace{-8pt}
\textbf{Context-Aware and Adaptive Automation}: A defining strength of NSE, as envisioned, is its ability to incorporate adaptive and context-sensitive intelligence into software engineering automation. Unlike purely deep learning-based approaches, which, while efficient in executing specific tasks, often require extensive retraining to adapt to new contexts, NSE leverages symbolic reasoning combined with controlled chaos to dynamically adjust to evolving requirements, environments, and software development paradigms. 

\vspace{-10pt}
\subsubsection{Challenges and limitations} \scalebox{1.2}{\faExclamationTriangle} \\

\vspace{-8pt}
\textbf{Handling randomness effectively}:
Controlling and managing randomness to achieve the desired level of variability without compromising reliability is a challenge. The predictions of neural models can lead to unwanted behaviors that can impact the consistency and dependability of automated software engineering tasks.\\

\vspace{-8pt}
\textbf{Encoding knowledge in symbolic learning systems}: 
Symbolic AI depends on explicitly encoded rules and logic, which require domain expertise to define.
Manually encoding symbolic constraints can be labor intensive and may not scale efficiently across diverse SE tasks. Ensuring that symbolic representations remain up-to-date as software engineering practices evolve is a nontrivial challenge. \\

\vspace{-8pt}
\textbf{Generalization across SE-related practices}:
Despite NSE born as a generalized paradigm based on NsPC by Velasco \etal \cite{velasco2025toward}, ensuring that models generalize effectively across different tasks and software development environments is critical and time-consuming. In this regard,  domain-specific optimizations may be required, making it harder to develop a one-size-fits-all NSE solution.\\

\vspace{-8pt}
\textbf{Deployment challenges}: 
The absence of well-established frameworks, tools, and benchmarks for NSE poses significant challenges for practitioners seeking to adopt and integrate it into existing software engineering workflows. Unlike traditional DL-based systems or symbolic approaches, NSE requires a hybrid infrastructure that also contemplates chaos, via deep learning non-determinism. This practical challenge can hinder the practical deployment of NSE-based solutions that would otherwise require interoperability with existing SE-tools, frameworks, and workflows. Bridging this gap is vital, as a lack of standardized support and tools would slow adoption, increase integration complexity, and limit the scalability of NSE-based systems. \\

\vspace{-8pt}
\textbf{Keeping Pace with Rapidly Evolving Software Engineering Practices}: Software engineering is a rapidly evolving discipline, and automation approaches designed for current practices risk becoming obsolete as new methodologies, frameworks, and technologies emerge. Ensuring that NSE remains adaptable and resilient to these continuous advancements is crucial for its long-term viability and requires joint effort provided by academics and industrials. \\

\vspace{-8pt}
\textbf{Ensuring Explainability Without Reducing Performance}: Increasing explainability often introduces additional computational complexity that can negatively impact performance. For example, while neural models rely on fast probabilistic approximations, symbolic reasoning methods on the other hand, may requires explicit rule processing, constraint solving, or logic-based inference, which can slow down execution.


\section{Conclusion and Future Directions}

\rev{Traditional probabilistic methods or chaos theory alone, while capable of modeling uncertainty, struggle to keep pace with the scale and adaptability of LLMs--that have fundamentally reshaped the way automation is approached in software engineering. For example, LLMs’ capability to generate context-aware suggestions in real-time requires handling the nondeterministic nature of the model, which exceeds the limitations of earlier approaches based solely on statistical inference or chaotic modeling.
The NSE framework addresses these challenges by integrating symbolic reasoning with neural learning and chaos-driven adaptability. The symbolic component ensures that the results of LLMs are interpretable and verifiable, enforcing correctness and providing structured inference. Meanwhile, the chaos-driven component simulates real-world uncertainty, helping LLMs adapt to changing environments and unseen data. 
This hybrid approach preserves the flexibility and scalability of LLMs while also ensuring their reliability and transparency in the face of dynamic, unpredictable software engineering tasks that increasingly rely on seamless collaboration between humans and intelligent systems to achieve the desired goals.
}


Looking ahead, we envision NSE as a key component of the future of software engineering automation, thanks to its ability to integrate controlled uncertainty. \rev{By adopting this hybrid approach, NSE facilitates more dependable, sustainable, and flexible automation solutions to address the complexities underlying modern software engineering practices that are blending with LLMs in increasingly dynamic, data-rich, and non-deterministic development environments. This integration enables systems not only to perform traditional tasks such as code generation and bug fixing, but also to adapt intelligently to changing requirements, ambiguous input, and evolving software ecosystems.}

\section*{Acknowledgment}
Dr. Poshyvanyk acknowledges the support of the National Science Foundation (NSF) under grant CCF-2311469.
\newpage
\bibliographystyle{ACM-Reference-Format}
\bibliography{main}

\end{document}